\documentclass[twocolumn,letterpaper,aps,prc,superscriptaddress,nofootinbib,showpacs,floatfix]{revtex4-1}
\usepackage{graphicx}
\usepackage{comment}
\usepackage{setspace}
\usepackage{amsmath}
\usepackage{amssymb}
\usepackage[textwidth=16cm,textheight=22cm]{geometry}
\usepackage{color}
\usepackage{indentfirst}
\usepackage{xspace}
\usepackage{hyperref}
\usepackage{verbatim}
\usepackage{epstopdf}

\begin{document}
\title{Charged-particle multiplicity and transverse-energy
  distribution using the Weibull-Glauber approach in heavy-ion collisions}

\author{Nirbhay~K.~Behera}
\email{nbehera@cern.ch}
\affiliation{Inha University, 100, Inharo, Nam-gu, Incheon 22212, Korea}

\author{Sadhana~Dash}
\email{sadhana@phy.iitb.ac.in}
\author{Bharati~Naik}
 \author{Basanta.~K.~Nandi}
\affiliation{Indian Institute of Technology Bombay, Mumbai 400076, India}
\author{Tanmay~Pani}
\affiliation{National Institute of Science Education and Research, Bhubaneswar 752050, India}

\date{\today}

\begin{abstract}
The charged-particle multiplicity distribution and the transverse-energy distribution measured in heavy-ion collisions at top RHIC and LHC energies are described using the two-component model approach based on a convolution of the Monte Carlo Glauber model with the Weibull model for particle production. The model successfully describes the multiplicity and transverse-energy distribution of minimum-bias collision data for a wide range of energies. The Weibull-Glauber model can be used to determine the centrality classes in heavy-ion collisions as an alternative to the conventional negative binomial distribution (NBD)-Glauber approach.
\end{abstract}

\maketitle

\section{Introduction}
The multiplicity distribution of charged particles emitted in
ultrarelativistic nucleus-nucleus collisions constitutes an important
global observable, which has been widely measured and studied to
understand the properties of the hot and dense matter created in such
collisions. Similarly, the transverse-energy ($E_{T}$) distribution is
also considered an additional global observable well suited for
probing the QCD medium. Both observables are connected to the
collision geometry, the entropy, and the initial energy density of the
system created in heavy-ion collisions \cite{xin,wong,bjroken}. 
Various experimental measurements suggest that charged-particle
pseudorapidity density can be successfully used to study the expansion
dynamics by Landau hydrodynamics
\cite{wong,landau,review,sarkisyan,stein}. Alternatively, the
transverse-energy measurements can also be used to probe the
longitudinal expansion of the produced system at midrapidity \cite{bjroken,review}. These observations establish a strong correlation between the two and their measurements provide significant constraints on the collision dynamics. The experimental observation of a linear correlation between the mean charged-particle multiplicity and the mean transverse energy further suggests the apparent equivalence. Hence, both of these global observables are usually the first measurements carried out in heavy-ion collision experiments to understand the underlying mechanism of particle production in heavy-ion collisions. In view of this, it is crucial to understand the charged-particle multiplicity and the transverse-energy distribution measured in such collisions.
\par
It is customary to understand the particle production mechanism in
nucleon-nucleon collisions before exploring complex systems like
nucleus-nucleus collisions. Recently, the Weibull model of particle
production based on the fragmentation and sequential branching
mechanism has been quite successful in describing the overall features
of multiparticle production in hadronic and leptonic collisions
\cite{weib1,weib2,weib3}. Previously, it was emphasized that the main
features of multiplicity distribution in nucleus-nucleus collisions
were just a consequence of the initial geometry of the collisions. The
interaction volume depends on the impact parameter (distance between
the centers of two colliding nuclei at the closest approach) of the
colliding nuclei and therefore on the number of participating nucleons
in that volume. However, it is impossible to measure the impact
parameter of the collision directly in experiments. Experimentally,
these quantities are best estimated by measuring the number of charged
particles produced or the transverse energy of the produced particles
in the collision. To establish a correspondence between these geometrical
quantities, like impact parameter, initial overlap volume, etc., with
measured charged-particle multiplicity, a Monte Carlo (MC)-based
Glauber model is widely used. In the MC Glauber model
\cite{gMC1,gMC2}, a nuclear collision (in pA and AA systems) is
modeled as a superposition of individual nucleon-nucleon interactions
\cite{glauber}. The initial overlap volume of two colliding
nuclei can be expressed in terms of the number of wounded
nucleons (nucleons which have undergone one or more binary
collisions). The number of such nucleons is known as the
number of participant nucleons, $N_{part}$.  The number of binary collisions
($N_{coll}$) among the nucleons depends on inelastic nucleon-nucleon cross section
($\sigma_{NN}^{inel}$). The MC Glauber model can be used to calculate
both these quantities for a Woods-Saxon type of initial nuclear
density distribution at a given value of the impact parameter. This
geometrical approach helps to provide a consistent description of
nuclear collisions in different systems ( p-A, d-A, and A-A )
when comparing data from different experiments to theoretical calculations \cite{alice1,alice2,phobos,star,phenix}. 

In this work, the two-component approach based on the Glauber model,
combined with the Weibull model of multiparticle production in
hadronic interactions, is implemented to generate the multiplicity
distribution in heavy-ion collisions. The obtained distribution is
compared with the experimental data measured at RHIC and LHC
energies. The Weibull-Glauber approach is similar to the procedure
adopted by the ALICE experiment for the centrality determination in
Pb-Pb collisions at 2.76 and 5.02 TeV using the negative binomial
distribution (NBD) \cite{alice1,alice2}. The transverse-energy
distribution in heavy-ion collisions was also described by using the
same formalism. An alternative method has been proposed to determine
the centrality in heavy-ion collisions using the model and HIJING
event generator \cite{twocompTheory1}.

\section{The Model}
The simple model aims to describe the qualitative features of the
charged-particle multiplicity and transverse-energy distribution in
heavy-ion collisions. In a nucleus-nucleus collision, each of the
wounded constituents gives rise to an ``ancestor,'' which then
fragments into the final state hadrons. The MC Glauber model
\cite{gMC1,gMC2} is used to simulate the collision
process of two nuclei on an event-by-event basis. The position
of the nucleons inside the nucleus is determined by the nuclear
density function, modeled by the available functional forms (Fermi
function, Hulth\'{e}n form, Woods-Saxon, uniform, etc). The impact
parameter is chosen randomly and the maximum value of the same is
fixed to twice the radius of the nucleus. The collision of two nuclei
is treated as a sequence of individual and independent collisions of
the nucleons, where the nucleons travel undeflected in a straight
path. The inelastic cross section, $\sigma_{NN}^{inel}$, is treated as
independent of the number of collisions a nucleon underwent
previously. The value of $\sigma_{NN}^{inel}$  is given as an input for the
MC Glauber model, which depends on the collision energy as used by
various experiments \cite{gMC1,alice1,star}. The model provides the
number of participants, $N_{part}$, and the number of binary
collisions, $N_{coll}$, for an event with a given impact parameter
and collision energy. To determine the particle multiplicity for a
single event, one defines the number of independent particle-emitting
sources, also known as ancestors \cite{alice1}. The number of
ancestors can be parametrized by assuming a suitable dependence on
$N_{part}$ and $N_{coll}$,  as the final multiplicity of an event
depends on the impact parameter of the collision. The event
multiplicity (or transverse energy) is expected to scale with
$N_{part}$ where the particle production is dominated by soft
processes, while the $N_{coll}$ scaling is observed where hard
processes dominate over soft particle production
\cite{twocompTheory2,twocomp1}.  A two-component approach has
been assumed \cite{alice1} where the number of ancestors have been
parametrized in terms of both $N_{part}$  and $N_{coll}$ as the following:
\begin{equation}
  N_{ancestors}  =  xN_{part} + (1-x)N_{coll}.
\label{ancestordep}
\end{equation}  

The expressed dependence takes care of the relative contribution
($x$) of both hard and soft processes in the final multiplicity. The
approach as described by Eq.\ref{ancestordep} in convolution with the
NBD has been very successful in describing the charged-particle
multiplicity densities at RHIC and LHC energies \cite{alice1}. But in
the present model, the charged-particle multiplicity per
nucleon-nucleon collision is parametrized by the Weibull
distribution. This latter assumption is motivated by the fact that in
minimum-bias $pp$ ($p\bar{p}$) collisions, the charged-particle
multiplicity distribution is nicely described by the Weibull function
for a wide range of energies \cite{weib1, weib2}. Therefore, one can
use the Weibull distribution as the statistical model of particle
production in nucleon-nucleon collision, with the Glauber model of
nucleus-nucleus collision to simulate the multiplicity distribution in
heavy-ion collisions.

The probability of producing $n$ particles per ancestor is given by the two-parameter Weibull distribution

\begin{equation}
  P(n;\lambda, k ) = \frac{k}{\lambda} \left( \frac{n}{\lambda}\right)^{k-1}~e^{- (\frac{n}{\lambda})^k}~,
\end{equation}
where $\lambda$ is related to the mean multiplicity per ancestor and
$k$ is related to the dynamics of particle production. For each event
generated by the Glauber model, the Weibull distribution is randomly
sampled $N_{ancestors}$ times to obtain the  amplitude of the number
of particles produced in that event. The process of
obtaining the multiplicity (or transverse-energy) distribution is
repeated for a large pool of events and for different values of
$\lambda$, $k$ and the two-component parameter
$x$. This is done to simulate an experimental 
multiplicity (or transverse-energy) distribution, which can be compared 
with measured experimental data. A $\chi$-square minimization method is
employed to obtain the best values of free parameters ($\lambda$, $k$
and $x$) for which the simulated distribution has good agreement with
the measured one. Alternatively, the goodness of the agreement was
also cross-checked by obtaining the ratio of measured amplitude to
that of the generated one. The obtained values of $\lambda$,
$k$ and $x$ for different collision systems and energies from the
Weibull-Glauber model are tabulated in Table \ref{paraTable}.

\begin{table*}[t]
\centering
\caption{The values of $\lambda$,
$k$ and $x$ obtained from the Weibull-Glauber model to describe the
minimum-bias charged-particle multiplicity and transverse-energy
distributions measured in different collision systems at RHIC and LHC energies.} 
\label{paraTable}
 \begin{tabular} {|p{2cm}|p{2cm}|p{4cm}|p{2.cm}|p{2.cm}|p{2.cm}|}
\hline
System & Energy & Distribution &$\lambda$ & $k$ & $x$ \\
\hline

Au$-$Au & 200 GeV & Charged particle  & 0.685 & 1.17 & 0.805 \\

Pb$-$Pb & 2.76 TeV & Charged particle & 33.20 & 1.17 & 0.8015 \\

Au$-$Au & 200 GeV & Transverse energy &4.58 & 1.17 & 0.802 \\

Pb$-$Pb & 2.76 TeV & Transverse energy &1.81 & 1.17 & 0.799\\
\hline

\end{tabular}
\end{table*}
\section{Results }
\begin{figure}
\includegraphics[width=7.8cm]{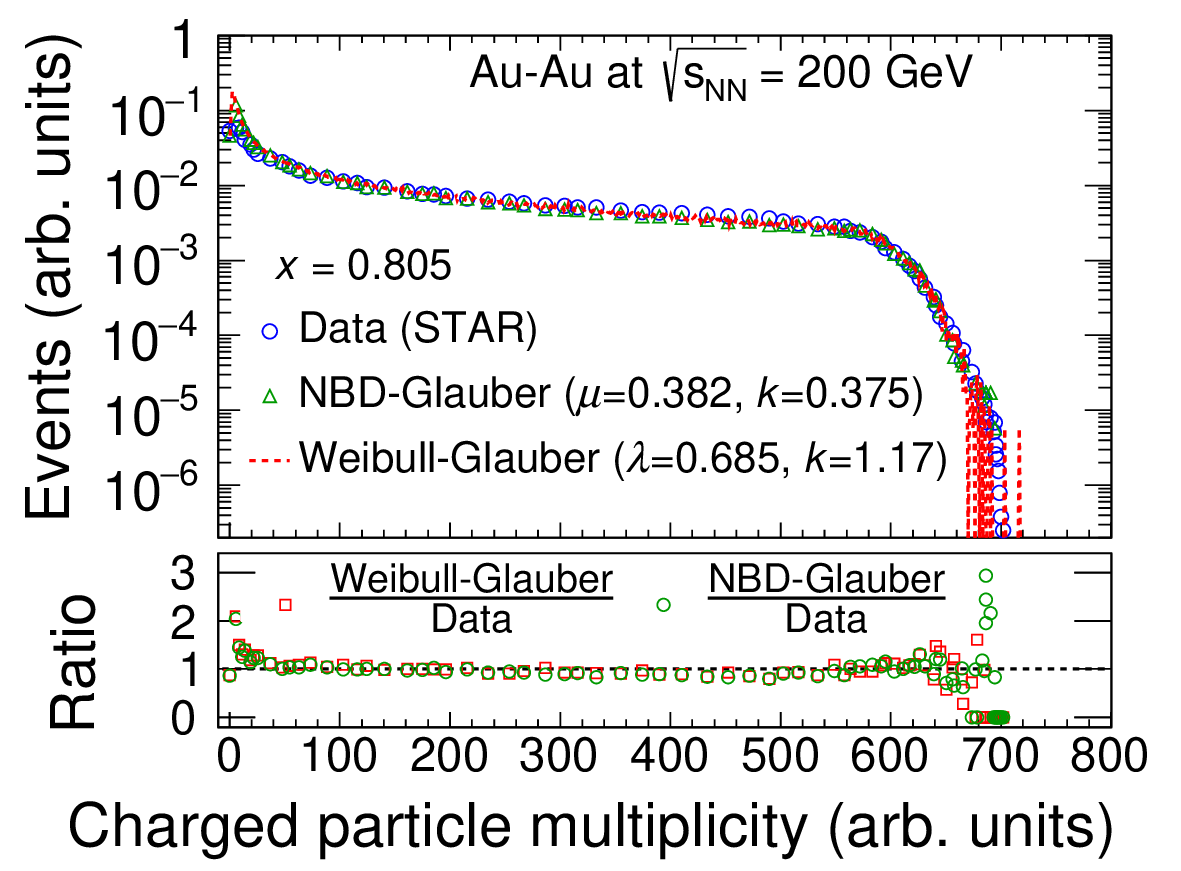}
\caption{(Color online) The charged-particle multiplicity distribution 
  measured by the STAR TPC detector in $|\eta| < 0.5$ for Au-Au
  collisions at $\sqrt{s_{NN}}$ = 200 GeV \cite{starMult}, represented by open circles. The distribution is compared with the Weibull-Glauber and NBD-Glauber models, which are shown by a dashed line and open triangles, respectively. The open squares (open circles) in the lower panel show the ratio between the measured multiplicity by the STAR experiment and the one obtained by the Weibull-Glauber (NBD-Glauber) model.} 
\label{midrapAuAu}
\end{figure}

\begin{figure}
\includegraphics[width=7.8cm]{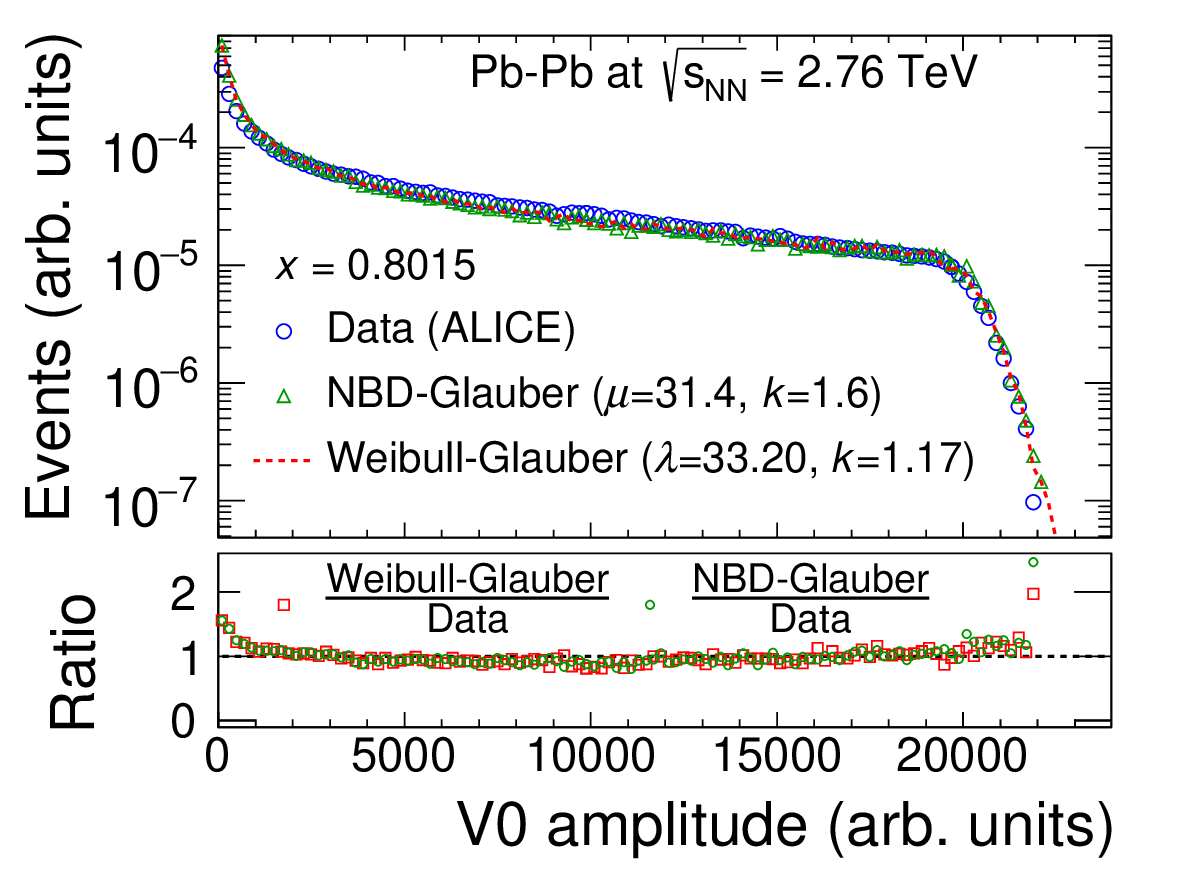}
\caption{(Color online) The distribution of the sum of amplitudes in 
  the VZERO scintillators in Pb-Pb collisions at $\sqrt{s_{NN}}$ =
  2.76 TeV \cite{alice1} shown by open circle. The distribution 
is compared with the Weibull-Glauber and NBD-Glauber model as
shown by dashed line and open triangles, respectively. The ratio of the
ALICE VZERO amplitude to that of the Weibull-Glauber (NBD-Glauber)
model is shown in the lower panel, which is represented by open
squares (open circles).}
\label{midrapPbPb}
\end{figure}

The method was used to describe the charged-particle multiplicity and
transverse-energy distribution measured by various experiments at RHIC
and LHC energies. The model is compared to the uncorrected
charged-particle multiplicity distribution measured in $|\eta| < 0.5$
in Au-Au collisions at $\sqrt{s_{NN}}$ = 200 GeV by STAR Time
Projection Chamber (TPC) detector, as shown in Figure \ref{midrapAuAu}
\cite{starMult}. The lower panel of Figure \ref{midrapAuAu} shows the ratio between the experimentally measured
multiplicity distribution and the one obtained from Weibull-Glauber
model. In ALICE, the amplitude of the VZERO detector is proportional to the charged-particle multiplicity. The ALICE VZERO detector consists of two scintillator arrays placed asymmetrically to the interaction point: VZERO-A ($2.8 < \eta < 5.1$)
and VZERO-C ($-3.7< \eta < 1.7$). The model was fitted to the VZERO
detector amplitude measured in Pb-Pb collisions at $\sqrt{s_{NN}}$ =
2.76 TeV by ALICE \cite{alice1}, which is shown in Figure
\ref{midrapPbPb}. The ratio between VZERO multiplicity and model
results are illustrated in the lower panel. It can be seen from Figure
\ref{midrapAuAu} and Figure \ref{midrapPbPb} that the
distributions obtained from the Weibull-Glauber model successfully describe the
charged-particle multiplicity distribution in Au-Au and Pb-Pb collisions at
$\sqrt{s_{NN}}$ = 200 GeV and 2.76 TeV, respectively. The model shows
some deviation from the charged-particle multiplicity distributions
for most peripheral and central events. This is because the most
peripheral events are contaminated by electromagnetic interactions and
by trigger inefficiency. The most central events are affected by
fluctuations of $N_{part}$ as well as by detector acceptance and
resolution effects \cite{alice1}, which have not been taken into
account in the model.
\par
\begin{figure}
\includegraphics[width=7.8cm]{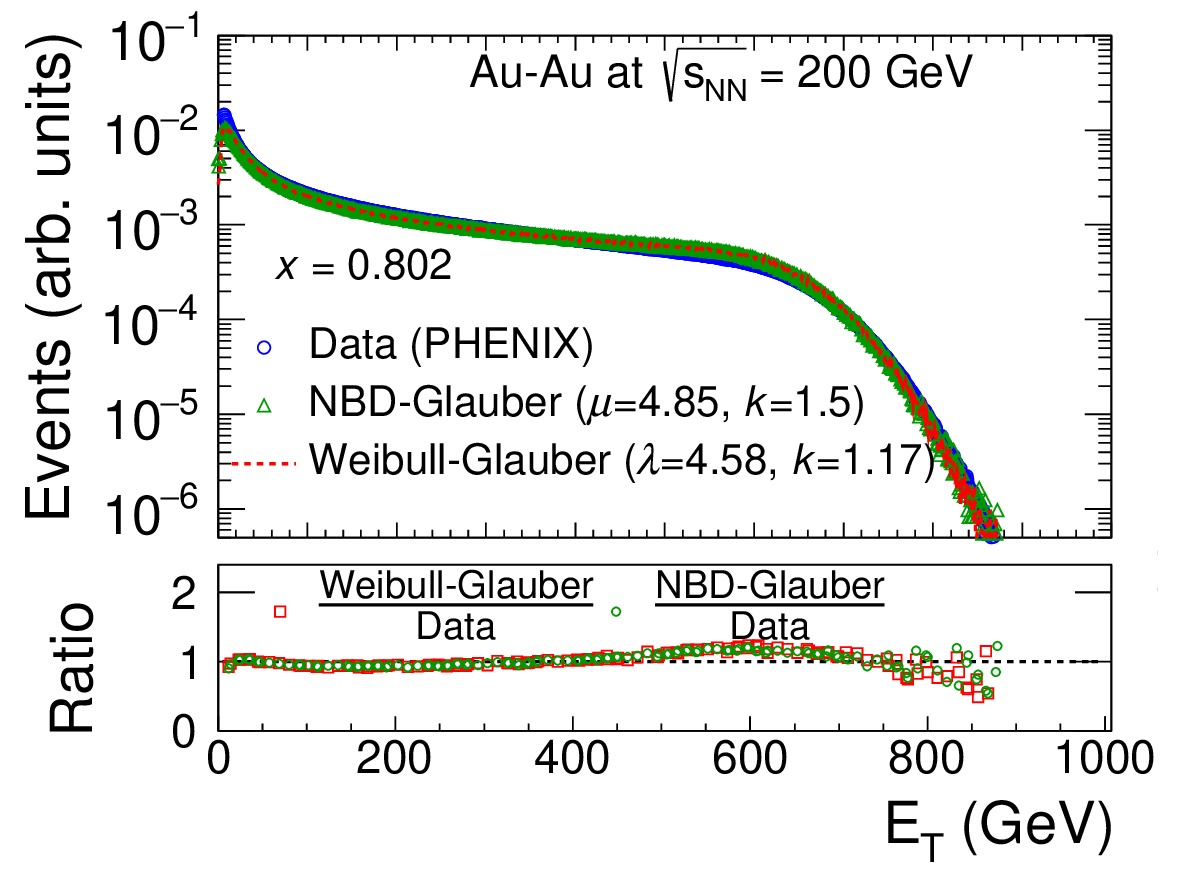}
\caption{(Color online) The open circle represents the transverse
  energy recorded in mid-rapidity
  by PHENIX lead-scintillator detector in minimum-bias Au-Au collision
  events at $\sqrt{s_{NN}}$ = 200 GeV \cite{phenixEt}. The data are
  compared with Weibull-Glauber and NBD-Glauber models, shown by a
  dashed line and open triangles, respectively. The open squares (open
  circles) in the lower panel represent the value of the ratio of
  experimentally measured transverse energy to the Weibull-Glauber
  (NBD-Glauber) model.}
\label{PhenixTransE}
\end{figure}
\begin{figure}
\includegraphics[width=7.8cm]{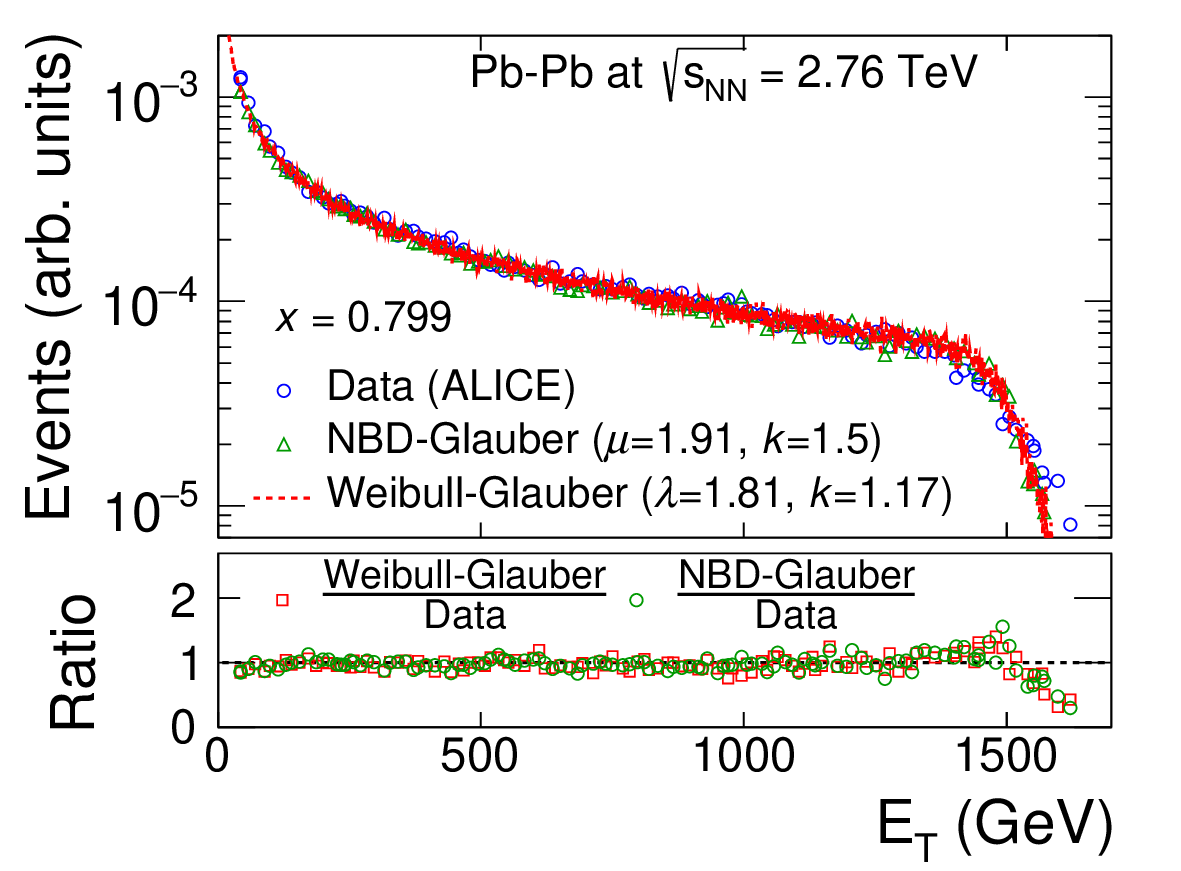}
\caption{(Color online) Mid-rapidity transverse-energy distribution of minimum-bias
  collision events in Pb-Pb collisions at $\sqrt{s_{NN}}$ = 2.76 TeV
  shown by open circles \cite{lhcet}. The dashed line (open triangles)
  represents the Weibull-Glauber (NBD-Glauber) model. The ratio of
  ALICE transverse-energy data and Weibull-Glauber (NBD-Glauber) model
  results are shown by open squares (open circles) in the lower
  panel.}
\label{AliceEt}
\end{figure}

The model was also used to fit the minimum-bias transverse-energy
distribution measured within the acceptance of $|\eta| < 1.0$ by
PHENIX experiment in Au-Au collisions at $\sqrt{s_{NN}}$ = 200 GeV
as shown in Figure \ref{PhenixTransE} \cite{phenixEt}. Recently, the
transverse energy distribution was also measured by ALICE experiment
for Pb-Pb collisions at $\sqrt{s_{NN}}$ = 2.76 TeV \cite{lhcet} at 
mid-rapidity ($|\eta| < 0.6$). Figure \ref{AliceEt} compares the $E_{T}$
distribution generated by the Weibull-Glauber model with the measured
distribution in Pb-Pb collisions at 2.76 TeV. The ratio of transverse-energy distribution obtained from experiments to the model
calculations is shown in the lower panels of Figure \ref{PhenixTransE}
and \ref{AliceEt}. Furthermore, all the distributions are also
compared with the estimations of the NBD-Glauber approach. It should
be noted that the value of the two-component $x$ parameter is the same
for both the approaches for a given distribution type and
energy. Thus, the two-component Weibull-Glauber approach provides a
nice description of the measured minimum-bias data for
charged-particle multiplicity and transverse-energy distribution, in
addition to the NBD-Glauber model.

\section{Determining Centrality Classes }
The experimental charged-particle multiplicity distribution can be
divided into different classes of geometrical collision by defining
sharp cuts in multiplicity or any other suitable detector variable to
define the same. These centrality classes also define the intervals of
hadronic cross section. It is very important to determine the correct
centrality class to study various physics observables as a function of
centrality or impact parameter and to compare the same at different
collision energies in heavy-ion experiments.

\begin{figure}
\includegraphics[width=7.8cm]{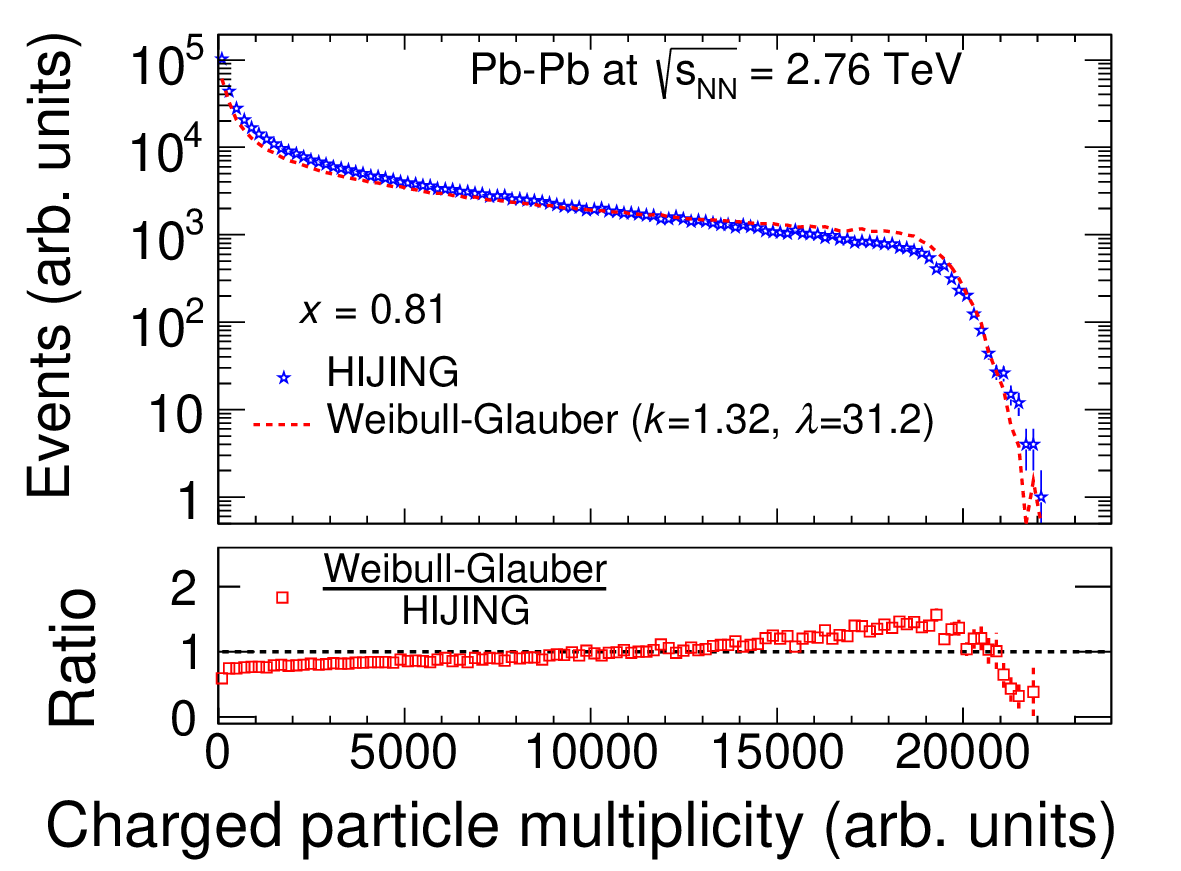}
\caption{(Color online) The charged particle distribution obtained
  from HIJING for Pb-Pb collisions at $\sqrt{s_{NN}}$ = 2.76 TeV is
  shown by open stars. The distribution is compared with the
  Weibull-Glauber model represented by the dashed line. The ratio
  between HIJING charged-particle multiplicity and the Weibull-Glauber
  model results is represented by open squares in the lower panel.}
\label{midrapHijing}
\end{figure}

\begin{figure}
\includegraphics[width=7.8cm]{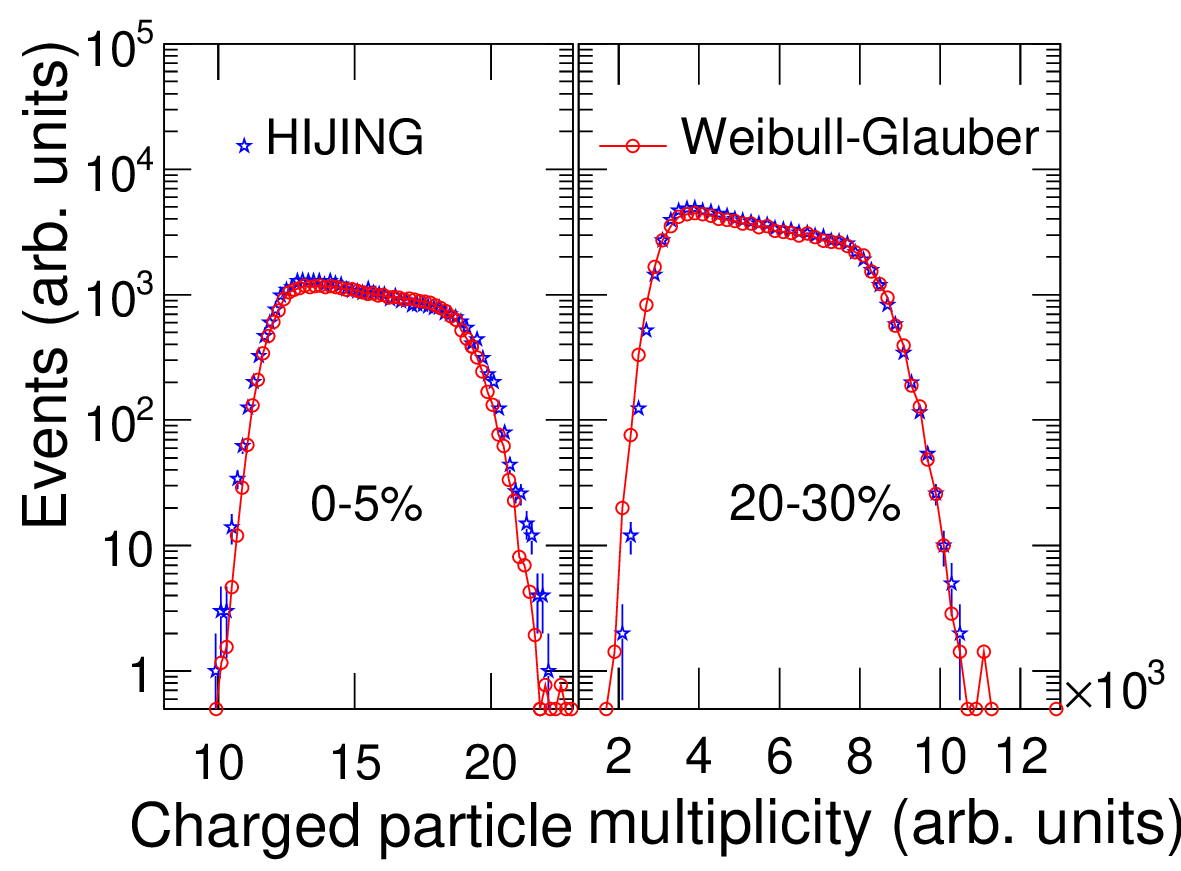}
\caption{(Color online) The charged particle distribution for
  different centralities obtained from HIJING for Pb-Pb collisions
  at $\sqrt{s_{NN}}$ = 2.76 TeV is shown by open stars. The centrality distributions are compared with the Weibull-Glauber model represented by open circles. The markers
are joined to guide the eye.}
\label{centHijing}
\end{figure}

To determine the centrality classes, the charged-particle multiplicity
distribution was obtained for Pb-Pb collisions at $\sqrt{s_{NN}}$ =
2.76 TeV, using HIJING event generator \cite{twocompTheory1}. The same
was simulated using the Weibull-Glauber model. Figure
\ref{midrapHijing} shows the comparison of the multiplicity
distribution obtained from HIJING with the one simulated with the
Weibull-Glauber approach. This creates a relationship between an
experimental observable and the phenomenological model of particle
multiplicity in nucleus-nucleus collisions using the Weibull-Glauber
approach. Thus, a given centrality class, defined by sharp cuts in the
geometrical properties [like $N_{part}$, $N_{coll}$, impact
parameter($b$), etc.] in the HIJING multiplicity distribution
corresponds to the same centrality class obtained using the
Weibull-Glauber model. The HIJING distribution was divided into
well-defined centrality intervals, as per the ratio of the area under
the curve in the selected interval to the total area under the
curve. The lower and higher limits of the multiplicity in a particular
centrality class are mapped to obtain the corresponding values of
impact parameter of the collision in the HIJING distribution. Using
the obtained impact parameter values from the HIJING model, the
multiplicity distribution is generated by the Weibull-Glauber approach
in that range of impact parameter. The values of $\langle N_{part}
\rangle$  and  $\langle N_{coll} \rangle$ obtained from the model are
then compared with the previously retained values from HIJING. This is
shown in Table \ref{NpartTable}. One can observe that the values
obtained from the model and HIJING are in close agreement with each
other. Figure \ref{centHijing} shows the good agreement between the
multiplicity distribution obtained from HIJING and the one simulated
with the Weibull-Glauber approach for two different centrality
classes. Hence, this method serves as an alternative way to determine
the approximate value of $\langle N_{part} \rangle$ and $\langle
N_{coll} \rangle$ for a given centrality class, for a given collision
system and energy.

\begin{table}[htb]
\centering
\caption{ Values  of $N_{part}$ from HIJING and Weibull-Glauber model for Pb$-$Pb collisions at  $\sqrt{s_{NN}}$ = 2.76 TeV} 
\label{NpartTable}
 \begin{tabular} {|p{1.5cm}|p{0.8cm}|p{0.8cm}|p{1.5cm}|p{1.25cm}| }
  \hline
Centrality (in \%) & $b_{min}$ (fm) & $b_{max}$ (fm) & $N_{part}$ (HIJING) & $N_{part}$ (Model) \\
\hline
 0 - 5\% & 0 & 5.0 & 364.7 & 354.9 \\ 
 20 - 30\% & 6.6 & 10.8 & 144.6 & 144.0 \\
 50 - 60\% & 11.0 & 13.8 & 34.01 & 34.24 \\
 70 - 80\% & 13.6 & 18.6 & 6.052 & 6.713 \\
 \hline
\end{tabular}
\end{table}

\section{Summary}

The charged-particle multiplicity and transverse-energy distributions in heavy-ion collisions at RHIC and LHC energies are well described by the two-component Glauber approach using the Weibull model of particle production. The Weibull-Glauber approach can be used to determine the centrality classes of heavy-ion collisions at a given energy. This is particularly significant regarding the applicability of the Weibull distribution to characterize the system created in heavy-ion collisions.

\begin{acknowledgments}
S.D., B.K.N., and B.N. would like to thank the Department of Science and Technology (DST), India, for supporting the present work. N.K.B. was supported by the National Research Foundation of Korea (NRF), Basic Science Research Program, funded by the Ministry of Education, Science and Technology (Grant No. NRF-2014R1A1A1008246).
\end{acknowledgments}

\end{document}